\documentclass[letterpaper,english,reprint, aps, superscriptaddress]{revtex4-2}
\usepackage[T1]{fontenc}
\usepackage[utf8]{inputenc}
\usepackage{amsmath}
\usepackage{amssymb}
\usepackage{graphicx}

\makeatletter

\pdfpageheight\paperheight
\pdfpagewidth\paperwidth

\usepackage{hyperref}
\usepackage{xcolor, soul} 
\usepackage[super]{nth}

\makeatother

\usepackage{babel}
\begin{document}
\title{Chiral Packings in Cylinders are Ultrasensitive to Confinement Deformation}
\author{Xuebin Wang}
\affiliation{School of Physics and Key Laboratory of Functional Polymer Materials
of Ministry of Education, Nankai University, and Collaborative Innovation
Center of Chemical Science and Engineering, Tianjin 300071, China}
\author{Jiahao Guo}
\affiliation{School of Physics and Key Laboratory of Functional Polymer Materials
of Ministry of Education, Nankai University, and Collaborative Innovation
Center of Chemical Science and Engineering, Tianjin 300071, China}
\author{Yao Li}
\email{liyao@nankai.edu.cn}

\affiliation{School of Physics and Key Laboratory of Functional Polymer Materials
of Ministry of Education, Nankai University, and Collaborative Innovation
Center of Chemical Science and Engineering, Tianjin 300071, China}
\begin{abstract}
Sphere packings in circular cylinders have attracted substantial research
interest, among which the discovery of chiral helical structures is
the most iconic. However, recent experimental results on zebrafish
do not match the known packing structures in circular cylinders. To
account for the inherent imperfections of biological tubes, we take
elliptic cylinders as the canonical deformation of circular cylinders
and investigate the densest packings of hard spheres in them using
simulation, theory, and experiments. Starting from the chiral structures
in circular cylinders, we demonstrate that even a weak cross-sectional
deformation can trigger entirely new phases, including ones that either
eliminate global chirality or significantly complicate the chiral
structures. This reveals the significant effect of cylindrical anisotropy.
The new helical phases under anisotropic confinement remain chiral
and develop hierarchical periodic structures, which are difficult
to obtain by simulations but are predicted by our newly developed
theory for helical phases in elliptic cylinders. The theory also predicts
double oscillated-chain phases without chirality, which perfectly
match the simulations. Our work offers fresh insights into understanding
packings in anisotropic cylinders, which will help researchers to
design new materials and to understand many living systems.
\end{abstract}
\maketitle
\textit{Introduction.}---The packing problem \citep{Torquato2010} originates from Kepler's
conjecture, which has attracted centuries of study by mathematicians
and physicists, and Hales eventually completed a proof with computer
assistance \citep{HALES2017}. It is important to understand how the
properties of soft matter systems depend on the spatial distribution
of constituents, which relates to the particle packing. Additionally,
studies of both ordered periodic arrangements and disordered arrangements
often invoke sphere packings. The packings of particles provide insights
into the structures and self-assembly properties of condensed matter
systems, and have been extensively investigated in mathematics, physics,
biology, computer science, and other related fields \citep{Donev2004,VanHecke2010,Royall2024}.

In recent years, the densest packings in confined systems have garnered
significant attention, with notable progress achieved in cylindrical
confinement. Such structures are commonly termed columnar crystals.
For the densest packings of hard spheres in circular cylinders \citep{Pickett2000,Mughal2011,Mughal2012,Chan2011,Chan2013,Yamchi2015,Zarif2021,Fu2016,Fu2017,Chan2019},
when the diameter ratio satisfies $D/d\in\left(1\textrm{, }2.7013\right)$,
all spheres are in contact with the cylinder wall, and various chiral
helical structures emerge. In addition, investigations of soft spheres
confined in circular cylinders have also observed many chiral helical
structures, which vary as the pressure is adjusted \citep{Winkelmann2017,Mughal2018,Winkelmann2019,Mughal2023}.
These columnar crystal structures have been observed in various systems,
such as foams \citep{Hutzler2009,Tobin2011,Meagher2015}, colloids
\citep{Khlobystov2004,Troche2005,Legoas2011,Tymczenko2008,Lohr2010,Jiang2013,Irannezhad2023},
and polymers \citep{Yu2006,Dobriyal2009,Shi2013}, and have had many
applications \citep{Ma2021,Liu2023,Vila-Liarte2022}. The densest
packings of nonspherical particles, such as ellipsoids \citep{Jin2020,Jin2021}
and nanodumbbells \citep{Wan2022} confined within circular cylinders,
have been investigated, revealing new structures.

While packings in ideal cylinders are elegant and receive great successes,
real-world tubes—especially in biological systems—are often imperfect.
Recent studies \citep{Norman2018,Sorrell2022,Curcio2023,Curcio2024}
have demonstrated that the cellular packing structures in the zebrafish
notochord have no counterparts among packings in circular cylinders,
suggesting that the notochord is anisotropic. However, how non-circular
cylindrical confinement tunes the densest packings remains unclear.
Here, we examine the packings of hard spheres in elliptic cylinders
to explore the effects of cylindrical anisotropy on the densest packing
structures.

\begin{figure*}
	\begin{centering}
		\includegraphics[width=180mm]{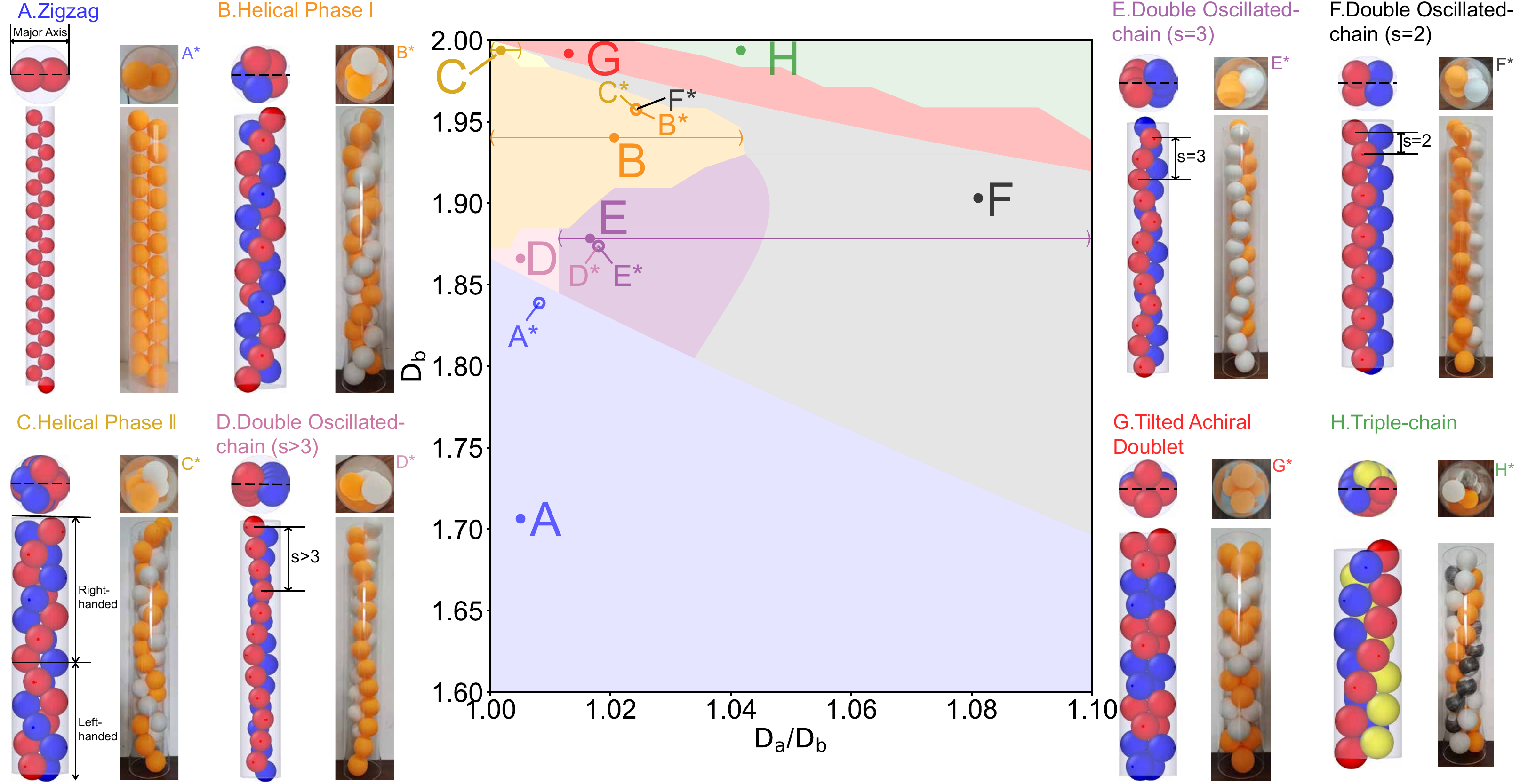}
		\par\end{centering}
	\caption{Phase diagram. Simulations of elliptic cylinders with minor axis $D_{b}\in\left[1.6\textrm{, }2\right]$,
		aspect ratio $D_{a}/D_{b}\in\left[1\textrm{, }1.10\right]$, and sphere
		numbers $N\le25$ are performed. Here, the sphere diameter is taken
		as the unit length, i.e., $d=1$, and all other quantities are expressed
		relative to $d$. The phase diagram contains eight regions of densest
		packing structures ($A-H$), with representative snapshots displayed
		beside it. Each snapshot is obtained at the parameters of its respective
		solid point in the phase diagram. The corresponding phase structures
		observed in the macroscopic experiments of table-tennis ball packings
		($A^{*}-H^{*}$) are shown beside the snapshots, which are obtained
		at the parameter values indicated by hollow points in the phase diagram
		(The points $G^{*}\approx\left(1.0086\textrm{, }2.03\right)$ and
		$H^{*}\approx\left(1.0072\textrm{, }2.08\right)$ are not marked).
		The spheres are colored differently in the snapshots and the experimental
		figures, which helps to better illustrate the structures. The three
		segments in the phase diagram correspond to the value ranges of the
		three subfigures in Fig. \ref{fig:3}.\label{fig:1}}
\end{figure*}

In this article, we study the densest packings of hard spheres confined
in elliptic cylinders, focusing on the small cylinder anisotropy regime.
Remarkably, we find that vanishingly small anisotropy is enough to
trigger the emergence of completely new phases. Helical phases are
tuned when compared with those observed in circular cylinders, whereas
double oscillated‑chain phases are absent in circular cylinders. We
develop a theory for helical phases in elliptic cylinders that, for
sufficiently large sphere numbers, approaches the ideal densest packings.
However, because hierarchical periodicity is disrupted at small sphere
numbers, the theoretical densest packings differ from the simulation
results. This theory is also employed for double oscillated-chain
phases, which yields results consistent with the simulations. All
phases predicted by our simulation and theory are found in our macroscopic
experiments on confined packings of table-tennis balls. Our results
encourage futher investigations into self-assembly processes in biological
systems featuring elliptic cylindrical structures.


\textit{Phase diagram.}---We investigate the densest packings of hard spheres in elliptic cylinders
with minor axis $D_{b}\in\left[1.60\textrm{,}2.00\right]$ and aspect
ratio $D_{a}/D_{b}\in\left[1.00\textrm{,}1.10\right]$. When $D_{a}/D_{b}=1.00$,
the cylinders have circular cross-sections, so for $D_{a}/D_{b}\leq1.10$,
the elliptic cylinders are obtained by slight deformations of the
circular cylinders. Through Monte Carlo (MC) simulations, we find
that even slight elliptic deformations lead to the emergence of various
densest packing structures (Fig. \ref{fig:1}): zigzag, helical phases,
double oscillated-chain phases, tilted achiral doublet, and triple-chain.
The most interesting results in this study are the first two findings:
the double oscillated-chain phases and the helical phases. These two
phases evolve from the known chiral helical phase observed in circular
cylinders. 

The double oscillated-chain configurations consist of two identical
chains with oscillated segments, which can be further classified by
the segment length $s$: $s=2$ (Fig. \ref{fig:1}F), $s=3$ (Fig.
\ref{fig:1}E), and $s>3$ (Fig. \ref{fig:1}D). These structures
are novel phases that are not observed in circular cylinders, and
they match well the packing structures of vacuolated cells in the
zebrafish notochord. They emerge when the cylinder is even slightly
deformed, indicating that the packings are highly sensitive to small
changes in cylinder shape. Therefore, they play an important role
in packing in slightly elliptic cylinders, which we will discuss in
detail later.

The helical phases evolve from the single/double helix structures
in circular cylinders and can be viewed as two helical chains intertwined
with each other. In the simulation, we observed two types of helical
phases: Helix phase $\mathrm{I}$ is monochiral, without any defects
(Fig. \ref{fig:1}B); helix phase $\mathrm{II}$ is bichiral, with
two defects that divide the structure into two segments of opposite
chirality (Fig. \ref{fig:1}C). However, due to limitations on the
number of spheres, the densest helical packings cannot be reliably
obtained by simulations, but we address this issue by developing a
new theory.

The latter three phases: the zigzag structure is identical to that
observed in circular cylinders (Fig. \ref{fig:1}A); the tilted achiral
doublet in elliptic cylinders is related to the achiral doublet in
circular cylinders, and its basic unit is formed by four spheres arranged
in a cross-shaped pattern (Fig. \ref{fig:1}G); the triple-chain structure
consists of three chains, in which spheres within each chain are not
necessarily in contact and do not form a helical structure (Fig. \ref{fig:1}H).
We also perform macroscopic experiments in which table-tennis balls
(with diameter $\sim40$mm) are packed into transparent elliptic tubes,
serving as a complement to our predictions. All phases predicted by
the simulations and theory are successfully identified in the experiments
by multiple packing trials with tubes of different sizes (Fig. \ref{fig:1}$A^{*}-H^{*}$).
Further experimental details are given in Section G of the Supplemental
Materials \citep{Supplemental}.

\textit{Zigzag phase.}---First, we discuss the simplest zigzag structure. For a circular cylinder
with the diameter equal to the sphere diameter ($D=d$), spheres form
a straight chain. When $D$ increases slightly beyond $d$, the available
space inside remains limited. Adjacent spheres move apart to reduce
vertical height differences, thereby maximizing space utilization
and forming a zigzag structure. Unlike circular cylinders, elliptic
cylinders provide more space along the major axis. Therefore, the
zigzag in elliptic cylinders strictly aligns with the major axis,
whereas in circular cylinders it can form along any diametrical direction
(details in Section A of the Supplemental Material\citep{Supplemental}).

\textit{Helical phases.}---To better explain the densest packing structures, we develop a theoretical
numerical method (see Methods). For a given number of spheres, we
can construct the packing structure using this method. Theoretically,
as the sphere number approaches infinity, this method yields the densest
structure; however, since only a finite number of spheres can be used
in calculations, periodic boundary conditions must be applied when
employing this method. In this way, when the sphere number $N\leq N_{max}$
(where $N_{max}$ is a positive integer), we can find the structure
with the maximum packing fraction among these cases (the corresponding
sphere number is the optimal number of spheres $N^{*}$). When $N_{max}$
is relatively small, we can calculate the results under a finite sphere
number and validate them against simulations. As $N_{max}$ increases,
we can approximately find the period of the structure and obtain the
densest helical packing. Further details are provided in Section C
of the Supplemental Material \citep{Supplemental}.

\begin{figure}
	\begin{centering}
		\includegraphics[width=88mm]{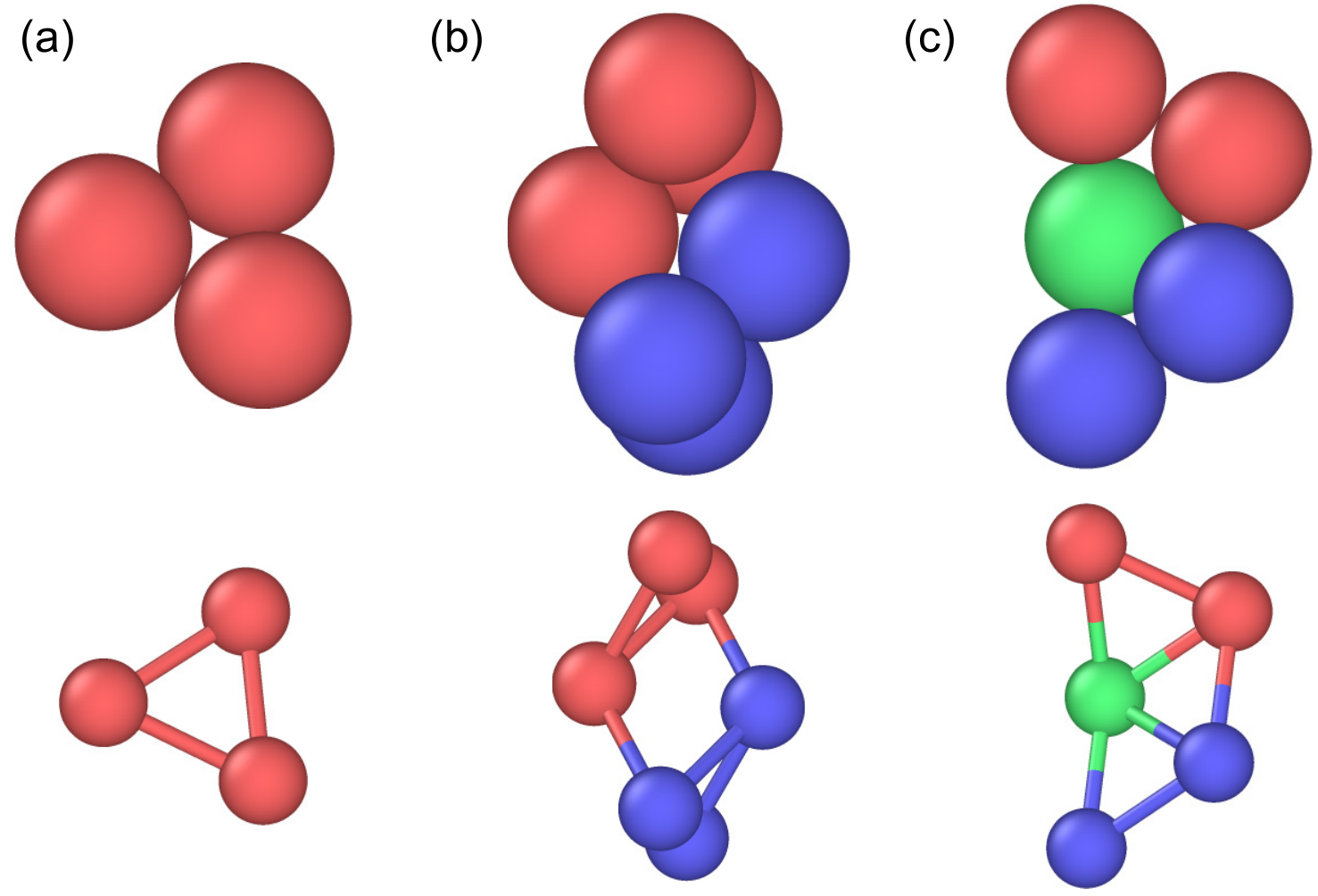}
		\par\end{centering}
	\caption{Typical blocks in packing structures. The first row shows snapshots
		of the actual spheres, and the bonds in the second row indicate the
		contact relationships between spheres. (a) A triplet of spheres comprises
		three spheres in contact with each other and with the wall, exhibiting
		chirality. It appears in various structures and is used in the triplet-sphere
		packing block method. (b) The central red and blue spheres have an
		angular difference of $\pi$ and do not contact each other, forming
		a defect that appears in helical phase $\mathrm{II}$. Triplets composed
		of red and blue spheres exhibit opposite chiralities. (c) Red-green
		and blue-green triplets display opposite chiralities, a feature that
		appears in double oscillated-chain phases.\label{fig:2}}
\end{figure}

\begin{figure}
	\begin{centering}
		\includegraphics[width=88mm]{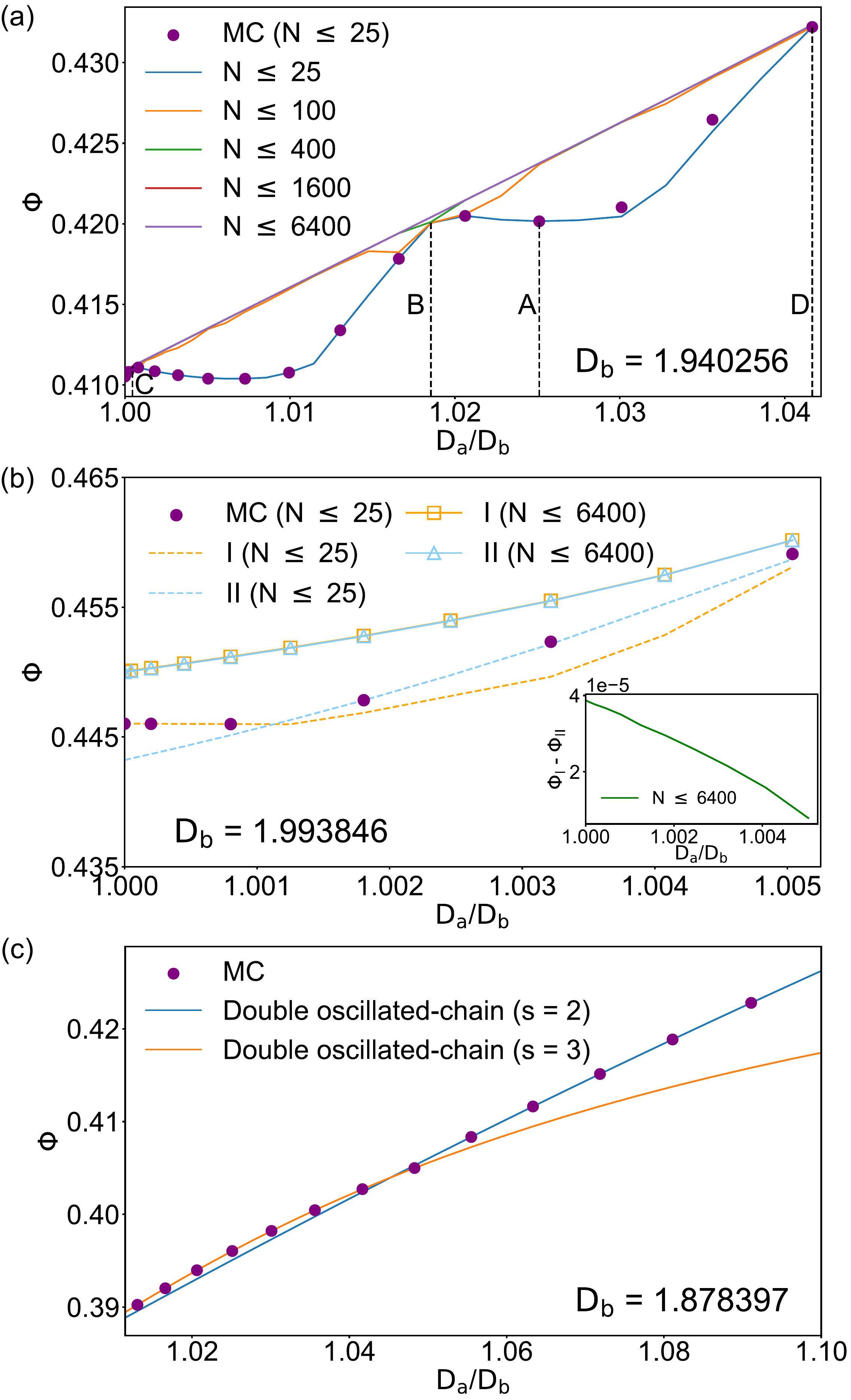}
		\par\end{centering}
	\caption{Packing fractions. The packing fractions $\phi$ from theoretical
		predictions (lines) are compared with the packing fraction $\phi$
		from $N\protect\leq25$ MC simulations (purple points). (a) For $D_{b}=1.940256$,
		densest structures are derived using the triplet-sphere packing block
		method at $N\protect\leq25$, $100$, $400$, $1600$, $6400$. Here,
		the curves for $N\protect\leq1600$ and $N\protect\leq6400$ nearly
		coincide. Structures corresponding to the dashed lines A, B, C, and
		D are presented in Fig. \ref{fig:5}(a)\textendash (d), respectively.
		(b) For $D_{b}=1.993846$, densest structures are found using the
		triplet-sphere packing block method (helical phase $\mathrm{I}$)
		and the defective triplet-sphere packing block method (helical phase
		$\mathrm{II}$) at small ($N\protect\leq25$) and large ($N\protect\leq6400$)
		sphere numbers. For large sphere numbers ($N\protect\leq6400$), both
		methods yield nearly identical packing fractions, but the small inset
		at the lower right indicates that the triplet-sphere packing block
		method achieves a slightly higher $\phi$. (c) For $D_{b}=1.878397$,
		the theoretical results for the double oscillated-chain ($s=2$ and
		$s=3$) perfectly match the simulation data and the transition between
		phase structures.\label{fig:3}}
\end{figure}

Helical phase $\mathrm{I}$ is formed by triplets of spheres {[}shown
in Fig. \ref{fig:2}(a){]} with identical chirality. For adjacent
spheres labeled $n$ and $n+1$ in the structure, the vertical height
difference between them is denoted by $\Delta z_{n\textrm{,}n+1}$.
In circular cylinders, the single and double helices can be characterized
by one and two values of $\Delta z_{n\textrm{,}n+1}$, respectively
\citep{Chan2019}. However, for helical phase $\mathrm{I}$ in elliptic
cylinders, $\Delta z_{n\textrm{,}n+1}$ takes multiple values (more
than two), indicating that the structure is composed of various distinct
triplets of spheres and can be constructed using the triplet-sphere
packing block method. Fig. \ref{fig:3}(a) shows the packing fraction
$\phi$ as a function of aspect ratio $D_{a}/D_{b}$ when the minor
axis $D_{b}=1.940256$ (the orange segment in the phase diagram of
Fig. \ref{fig:1}). The triplet-sphere packing block method applied
for sphere numbers $N\leq25$ predicts most of the MC simulations.
Extending the method to larger $N$ ($N\le100$, $400$, $1600$,
and $6400$), the packing fraction $\phi$ increases with $N$ and
clearly exceeds the simulations. This implies that periodic boundary
conditions with finite $N$ affect the densest packings. When $N$
is small, we may not be able to obtain the structures of the densest
helical phase. To satisfy the periodic boundary conditions, several
spheres at the boundary do not form a triplet of spheres but instead
form defects, resulting in a low packing fraction. As $N$ increases,
the densest helical structures can possibly be reached. Meanwhile,
the defects caused by the periodic boundary conditions may vanish
or become less important, so the packing fraction increases significantly.
The curves for $N\leq6400$ and $N\leq1600$ are almost identical,
indicating that the ideal densest packing for the infinite $N$ system
is approached. For $N\leq25$, to satisfy the periodic boundary conditions,
the triplet-sphere packing block method must impose a large $\Delta z_{N\textrm{,}1}=\Delta z_{N\textrm{,}N+1}$,
which can be reduced in some cases when the simulated configurations
partially break the triplet-sphere structures, thereby yielding a
slightly higher $\phi$.

\begin{figure}
	\begin{centering}
		\includegraphics[width=88mm]{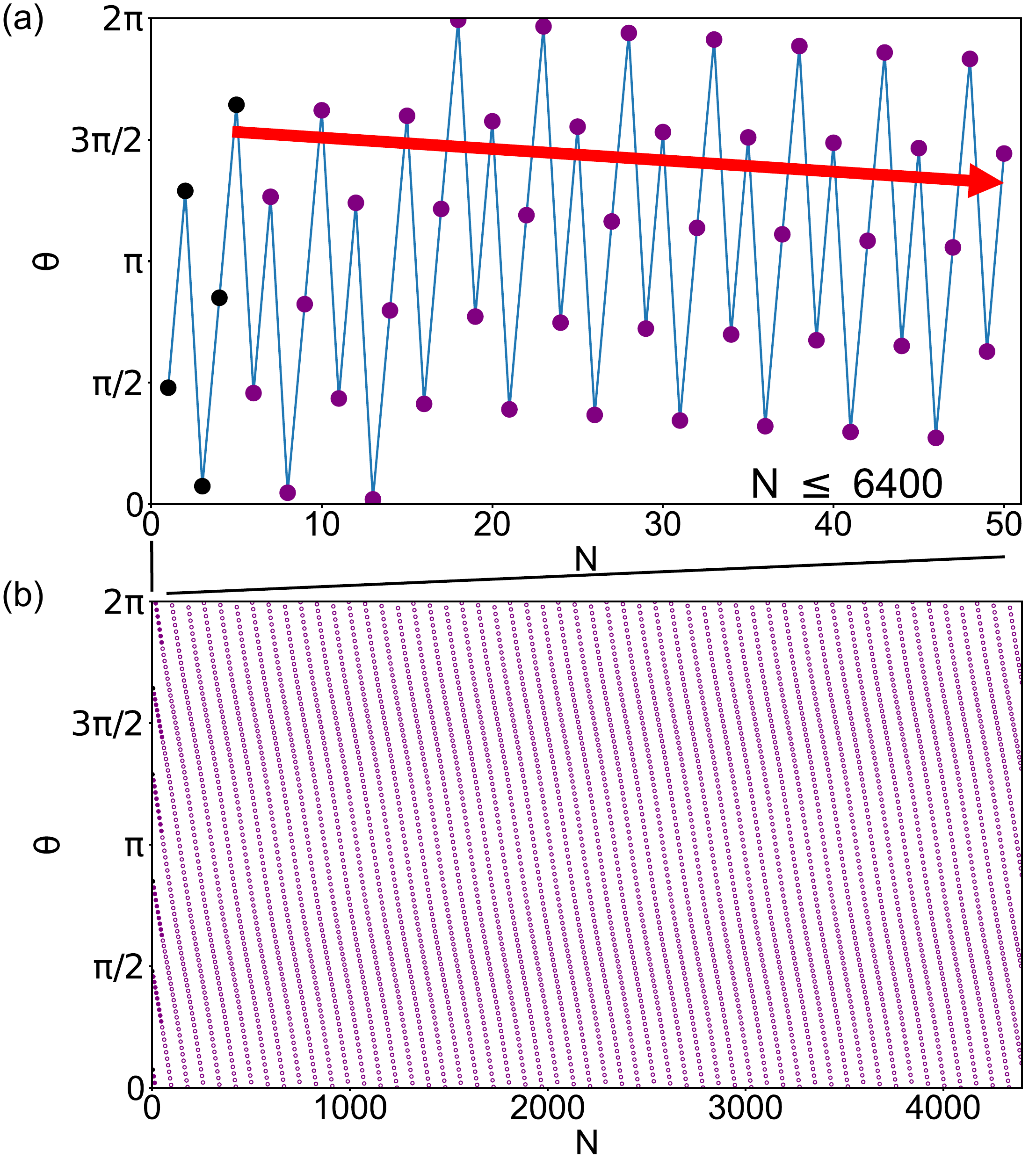}
		\par\end{centering}
	\caption{The hierarchical periodic structure. When $D_{b}=1.940256$ and $D_{a}/D_{b}=1.02512$,
		the optimal number of spheres is $N^{*}=4397$ for $N\protect\leq6400$.
		Solid dots denote the first 50 spheres, hollow dots denote the following
		spheres, and black solid dots mark the first subperiod. (a) The angular
		positions $\theta$ of the first 50 spheres are shown, revealing that
		the structure is composed of five-sphere subperiods. Red arrows indicate
		the angular offsets of spheres across subperiods. (b) The angular
		positions $\theta$ of all 4397 spheres reveal a longer-period structure.\label{fig:4}}
\end{figure}

\begin{figure*}
	\begin{centering}
		\includegraphics[width=180mm]{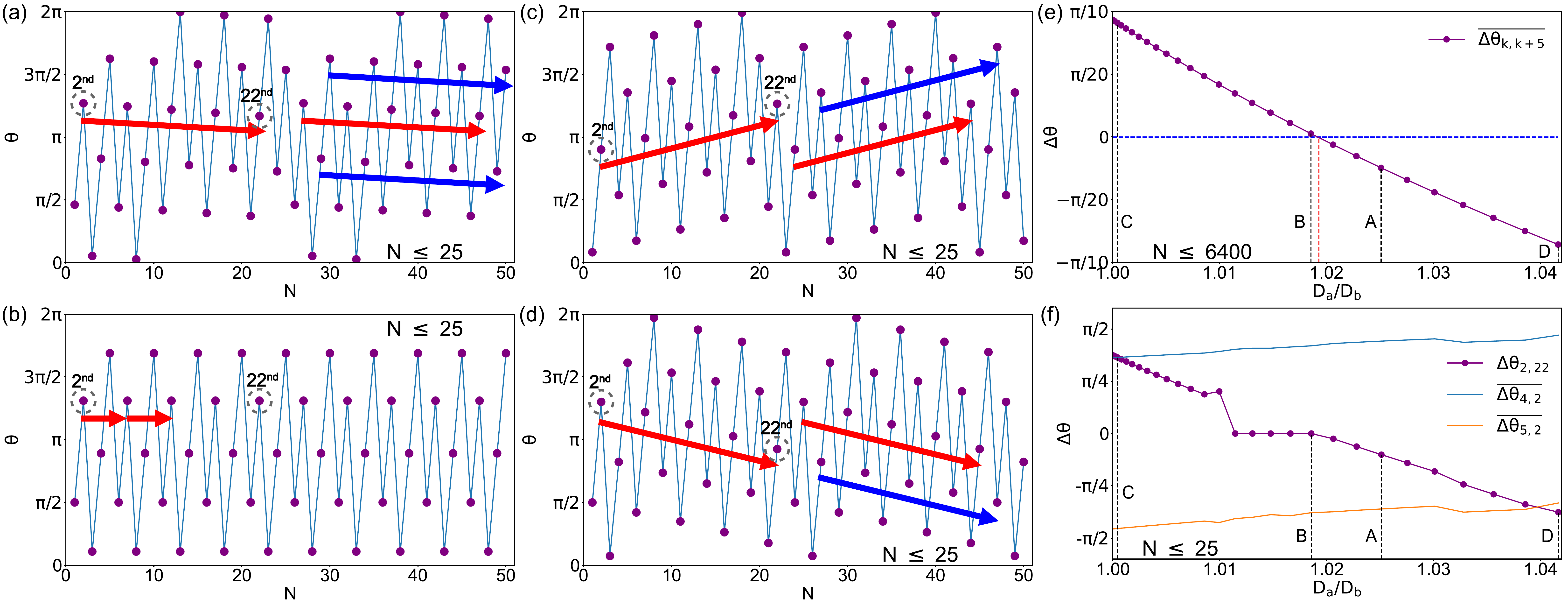}
		\par\end{centering}
	\caption{Discussion of the effect of periodic boundary conditions on the structure.
		For minor axis $D_{b}=1.940256$: for sphere number $N\protect\leq25$,
		we use the triplet-sphere packing block method to find the densest
		structures, and then apply periodic boundary conditions to obtain
		the angular positions of the first 50 spheres for four cases. (a)
		aspect ratio $D_{a}/D_{b}=1.02512$; the optimal number of spheres
		is $N^{*}=25$. (b) $D_{a}/D_{b}=1.018554$, $N^{*}=5$. (c) $D_{a}/D_{b}=1.000450$,
		$N^{*}=22$. (d) $D_{a}/D_{b}=1.041667$, $N^{*}=23$. The arrows
		indicate how the angular positions of spheres change across subperiods
		over one period. In Figure (a), the spheres marked by the left red
		arrow are not collinear with those marked by the three right arrows.
		In Figures (b), (c), and (d), the spheres marked by the left red arrow
		lie on the same line as the spheres marked by the red arrow (b), the
		upper blue arrow (c), and the lower blue arrow (d) on the right, respectively.
		(e) Using the triplet-sphere packing block method to find the theoretical
		densest packings ($N\protect\leq6400$), we find that the average-angular
		offset of the spheres between subperiods, $\overline{\Delta\theta_{k,k+5}}$,
		decreases as the aspect ratio increases, and that when the aspect
		ratio $D_{a}/D_{b}\approx1.019$ (marked by the red dashed line, which
		is near case B), the average-angular offset is zero. (f) For $N\protect\leq25$,
		we plot the angular difference $\Delta\theta_{2,22}$ between the
		$\nth{2}$ and $\nth{22}$ spheres of the constructed structures for
		various aspect ratios, and we further compute the average-angular
		differences between the $\nth{2}$ and $\nth{4}$ spheres, $\overline{\Delta\theta_{4,2}}$,
		and between the $\nth{2}$ and $\nth{5}$ spheres, $\overline{\Delta\theta_{5,2}}$,
		within the subperiods. Case C is located in the region where $\Delta\theta_{2,22}$
		approaches $\overline{\Delta\theta_{4,2}}$, while case D is located
		in the region where $\Delta\theta_{2,22}$ approaches $\overline{\Delta\theta_{5,2}}$.\label{fig:5}}
\end{figure*}

To further study the helical structures, we observe the angular position
of each sphere in the densest packings and discover the hierarchical
periodic structures. When $N\leq6400$, with minor axis $D_{b}=1.940256$
and aspect ratio $D_{a}/D_{b}=1.02512$, the optimal number of spheres
is $N^{*}=4397$, and the angular position $\theta$ of each sphere
is plotted in Fig. \ref{fig:4}. Fig. \ref{fig:4}(a) shows the angular
positions of the first 50 spheres. The red arrow indicates the angular
change between spheres separated by four other spheres. The direction
of each change is consistent, and the magnitudes of the changes are
similar. This suggests that a group of five spheres exhibits an approximate
periodicity and nearly completes one full rotation around the inner
wall of the elliptic cylinder. However, the rotation angle differs
slightly from $2\pi$, producing subperiods with offset angular positions.
Multiple subperiods combine to form a higher-level period. Fig. \ref{fig:4}(b)
presents the full period of all 4397 spheres, suggesting that the
helical phase forms a higher-level periodic structure composed of
nested subperiods.

In general, the packing fractions of the constructed structures for
$N\leq25$ match the simulation results but differ significantly from
the theoretical densest packings. For example, in Fig. \ref{fig:3}(a),
the packing fraction of the constructed structure for $N\leq25$ represented
by dashed line A is obviously lower than that of the theoretical densest
structure. The constructed structure represented by dashed line A
is given in Fig. \ref{fig:5}(a). Comparing the structure shown in
Fig. \ref{fig:5}(a) with the theoretical helical structure in Fig.
\ref{fig:4}(a) reveals that, for $N\leq25$, the angular positions
undergo a sudden change at the period boundary. This indicates that,
for structures constructed with a small number of spheres, periodic
boundary conditions have a significant influence, thereby preventing
the formation of the hierarchical period.

Specifically, for $N\leq25$, the constructed structures corresponding
to the cases indicated by dashed lines B, C, and D in Fig. \ref{fig:3}(a)
have packing fractions approaching the densest packings. Dashed lines
B, C, and D represent cases B, C, and D, and the corresponding structures
are shown in Figs. \ref{fig:5}(b), \ref{fig:5}(c), and \ref{fig:5}(d),
respectively. Fig. \ref{fig:5}(e) shows that the theoretical densest
structure for case B has a vanishing average-angular offset between
subperiods $\overline{\Delta\theta_{k,k+5}}\triangleq\overline{\theta_{k+5}-\theta_{k}}\approx0$.
Therefore, for $N\leq25$, the subperiod becomes the period, and the
constructed structure approximates the densest packing {[}see Fig.
\ref{fig:5}(b){]}. In Fig. \ref{fig:5}(f), we compare the angular
difference $\Delta\theta_{2,22}\triangleq\theta_{22}-\theta_{2}$
between the $\nth{2}$ and $\nth{22}$ spheres with the average-angular
differences $\overline{\Delta\theta_{4,2}}\triangleq\overline{\theta_{5k+2}-\theta_{5k+4}}$
and $\overline{\Delta\theta_{5,2}}\triangleq\overline{\theta_{5k+2}-\theta_{5k+5}}$
(the average-angular differences between the $\nth{2}$ and $\nth{4}$,
and between the $\nth{2}$ and $\nth{5}$ spheres within each subperiod,
respectively). The angular difference $\Delta\theta_{2,22}$ approaches
$\overline{\Delta\theta_{4,2}}$ near C, and approaches $\overline{\Delta\theta_{5,2}}$
near D. The structures for cases C and D are shown in Figs. \ref{fig:5}(c)
and \ref{fig:5}(d), respectively. In Fig. \ref{fig:5}(c), the sequence
formed by the $\nth{2}$ sphere within each subperiod of one period
is collinear with the sequence formed by the $\nth{4}$ sphere within
each subperiod of the subsequent period. Fig. \ref{fig:5}(d) shows
a similar collinear relationship between the sequences formed by the
$\nth{2}$ and $\nth{5}$ spheres. Collinearity reduces small-system
periodic-boundary effects, enabling the structure to approach the
theoretical densest packing fractions. More detailed analysis is given
in Section D of the Supplemental Material \citep{Supplemental}.

In helical phase $\mathrm{II}$, helical chains invert chirality twice
(Fig. \ref{fig:1}C), with defects emerging at each inversion point.
These defects resemble those introduced in the related work \citep{Yamchi2015,Zarif2021}.
The triplets of spheres above and below the defect exhibit opposite
chiralities, and the two spheres at the defect have an angular difference
of $\pi$ without contact {[}Fig. \ref{fig:2}(b){]}. The two defects
divide the structure into two segments of equal sphere numbers but
opposite chiralities, each defect acting as a center of inversion
symmetry. By placing one defect at the center, generating the other
at the periodic boundary through periodic boundary conditions, and
accounting for chirality reversal, the structure can be constructed
using the triplet-sphere packing block method. This process is referred
to as the defective triplet-sphere packing block method. Fig. \ref{fig:3}(b)
shows the case with minor axis $D_{b}=1.993846$ (the yellow segment
in the phase diagram of Fig. \ref{fig:1}), where increasing the aspect
ratio $D_{a}/D_{b}$ causes the structure to transition from helical
phase $\mathrm{I}$ to helical phase $\mathrm{II}$. For $N\leq25$,
structures built via the triplet-sphere packing block method exhibit
higher packing fraction $\phi$ at small $D_{a}/D_{b}$, while the
defective triplet-sphere packing block method achieves higher $\phi$
at large $D_{a}/D_{b}$, which is consistent with the trend observed
in the simulation results. However, for the defective triplet-sphere
packing block method, the resulting $\phi$ is significantly lower
than the simulation results at larger $D_{a}/D_{b}$. This is because
the structure achieves higher packing fraction by partially disrupting
some triplets of spheres, which may also be related to periodic boundary
conditions and the finite number of spheres. For large sphere numbers
{[}$N\le6400$ in Fig. \ref{fig:3}(b){]}, both construction methods
produce nearly identical packing fractions. In fact, the triplet-sphere
packing block method yields a slightly higher packing fraction $\phi$,
indicating that helical phase $\mathrm{II}$ transitions to helical
phase $\mathrm{I}$ as $N$ increases, and suggesting that helical
phase $\mathrm{II}$ is specific to small $N$.

\begin{figure}
	\begin{centering}
		\includegraphics[width=88mm]{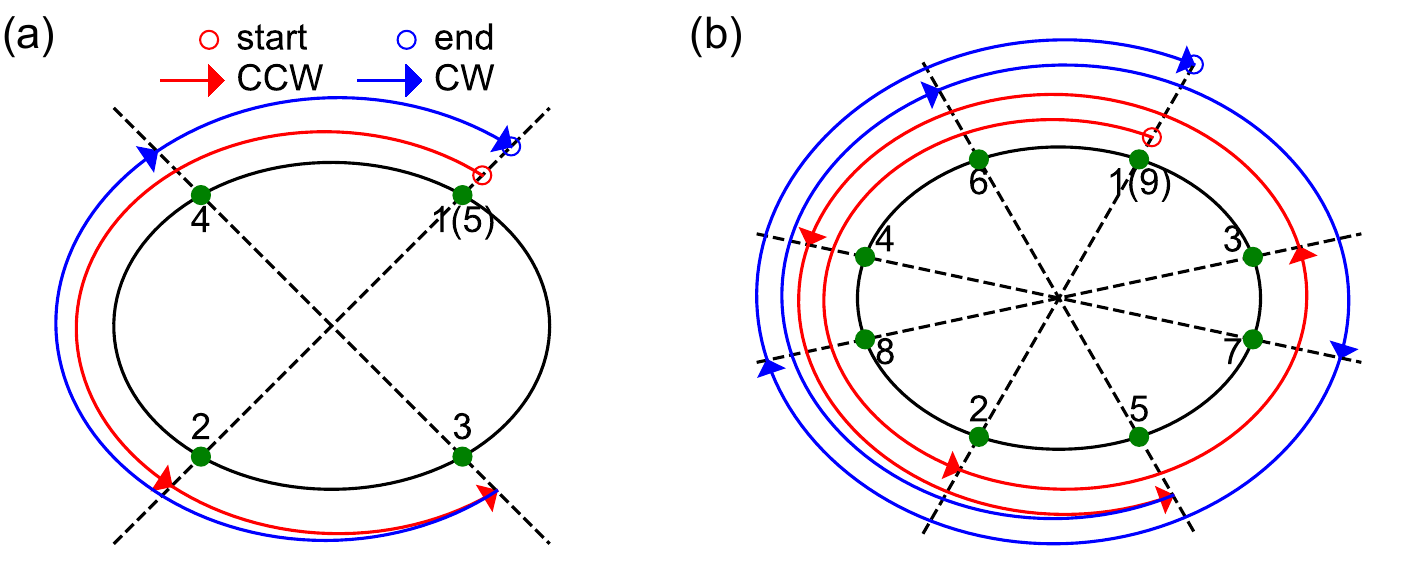}
		\par\end{centering}
	\caption{Sketches of the double oscillated-chain for segment lengths $s=2$
		and $s=3$ (see Fig. \ref{fig:1}). Spheres are labeled from bottom
		to top, and arrows indicate the rotational directions. Red arrows
		denote counterclockwise rotations, and blue arrows denote clockwise
		rotations. A single period of (a) the double oscillated-chain ($s=2$)
		and (b) the double oscillated-chain ($s=3$) is demonstrated. During
		packing, changes in rotational direction give rise to chirality reversals.\label{fig:6}}
\end{figure}

\textit{Double oscillated-chain phases.}---The slight anisotropy even produces the novel double oscillated-chain
structure, which also comprises triplets of spheres. However, the
chirality alternates between positive and negative, forming a periodic
pattern, which makes global structures achiral. Figs. \ref{fig:1}D,
\ref{fig:1}E, and \ref{fig:1}F show the phase structures, while
Fig. \ref{fig:2}(c) shows the details of the chirality transition
point. To elucidate the packing structures, top-view sketches of the
double oscillated-chain for $s=2$ and $s=3$ are provided in Fig.
\ref{fig:6}. In this figure, spheres are packed sequentially from
bottom to top and labeled accordingly, and arrows indicate the rotational
directions between adjacent spheres. Chirality reversals in the triplets
of spheres arise from changes in the rotational direction. The double
oscillated-chain ($s=2$) has a four-sphere periodicity, with all
sphere positions symmetric with respect to both the $x$- and $y$-axes.
The rotational direction from sphere 2 to sphere 3 is counterclockwise,
while that from sphere 4 to sphere 5 is clockwise, resulting in a
reversal of chirality. Notably, the angular difference between sphere 1
and sphere 2 is $\pi$, and that between sphere 3 and sphere 4 is
also $\pi$, which makes no distinction between clockwise and counterclockwise
rotations (one choice of rotational direction is shown in Fig. \ref{fig:6}).
The double oscillated-chain ($s=3$) has an eight-sphere periodicity,
with all sphere positions symmetric with respect to both axes. Analogous
analysis also reveals alternating rotational directions during packing.
The two structures can still be generated via the triplet-sphere packing
block method, provided that the packing permits reversal of rotational
directions. Specifically, for $s=2$, we place four spheres in the
cylinder. During the packing process, the rotations from sphere 1
to sphere 2 and from sphere 2 to sphere 3 have the same direction,
while the rotation from sphere 3 to sphere 4 is opposite that direction.
Similarly, for $s=3$, the rotation direction from sphere 1 to sphere
5 is opposite to that from sphere 5 to sphere 8. Varying the initial
sphere positions yields different packings, and the structure with
the maximum packing fraction is chosen as the theoretical result.
Owing to the exact periodicity, theoretical predictions match simulation
results perfectly, as shown in Fig. \ref{fig:3}(c) ($D_{b}=1.878397$,
the purple segment in the phase diagram of Fig. \ref{fig:1}). The
double oscillated-chain ($s>3$) follows analogous but more complex
patterns, which are not detailed here.

\textit{Other phases.}---When nearest-neighbor spheres lie at the same height, two adjacent
pairs form an intersecting cross structure, yielding the achiral doublet
(observed in circular cylinders at $D=2$). Increasing the minor axis
$D_{b}$ or the aspect ratio $D_{a}/D_{b}$ transforms the cross structure
from horizontal to tilted, with adjacent cross structures exhibiting
opposite tilt directions, termed the tilted achiral doublet. The triple-chain
structure comprises three distinct chains; however, these chains may
include segments where spheres are not in direct contact but still
exhibit a tendency to form a chain. Additionally, no perfect helical
structure is formed in this case. The tilted achiral doublet and the
triple-chain may be influenced by periodic boundary conditions and
finite sphere numbers, which will be the focus of our future work.

\textit{Discussion.}---In summary, we investigated the densest packings of hard spheres confined
within near-circular elliptic cylinders. We used simulated annealing
to study systems with minor axis $D_{b}\in\left[1.6\textrm{, }2\right]$,
aspect ratio $D_{a}/D_{b}\in\left[1\textrm{, }1.10\right]$, and sphere
numbers $N\leq25$, identifying several distinct phase configurations.
Through analysis of triplets of spheres, we elucidated the helical
phases and double oscillated-chain phases. The most iconic packings
in circular cylinders, the chiral helical structures, are sensitive
to the anisotropy of confinement. We observed that the helical phases
in elliptic cylinders consist of multiple triplets of spheres that
form complex hierarchical structures. The double oscillating-chain
phases are formed by packing triplets with alternating chirality,
resulting in the cancellation of the global chirality. It is merely
a slight change in the anisotropy of the cylinder that brings about
these changes. In circular cylinders, the densest structure transitions
from zigzag to helical phases as the diameter increases, whereas in
elliptic cylinders, the transition involves intermediate double oscillated‐chain
phases. This can be understood simply: when the minor axis is sufficiently
large to form a helix, increasing the aspect ratio provides more space
along the major axis, causing the spheres to separate toward both
ends. However, because the available space at the ends is limited,
a complete helical structure cannot be formed, forcing the spheres
into oscillatory arrangements.

This work reveals the surprising complexity that emerges from an impressively
simple system. Using elliptic tubes as the typical geometry, we show
that even infinitesimal anisotropy can lead to entirely new phases—an
effect we anticipate will be ubiquitous across a broad spectrum of
anisotropic confinements. The predictions have been realized by our
macroscopic experiments and are expected to encourage further experiments
across many length scales. The findings will inevitably stimulate
future explorations of richer interactions from atoms and nanoparticles
to biological morphogenesis and even macroscopic phenomena in anisotropic
confinements, and promise to yield more intriguing packing structures.
A compelling illustration is our ongoing project on water encapsulated
in elliptical carbon nanotubes, which already exhibits several of
the packing structures presented here. Our work gives a new way to
design materials through packings under confinement, and helps researchers
to understand many puzzles in living systems.

\textit{Methods.}---We set the sphere diameter to $d=1$ (unit length), with all other
quantities expressed relative to $d$. For an elliptic cylinder, the
major axis is denoted by $D_{a}$ and the minor axis by $D_{b}$;
the aspect ratio $D_{a}/D_{b}$ quantifies the elliptical cross-sectional
flattening. The cross-sectional plane lies in the $xy$-plane, and
the cylinder’s axis is aligned along the $z$-axis. Circular cylinders
necessitate periodic boundary conditions along the $z$-axis and twisted
periodic boundary conditions in the $xy$-plane, owing to their invariance
under arbitrary rotations about the central axis. In contrast, elliptic
cylinders are anisotropic in the $xy$-plane; thus, only periodic
boundary conditions along the $z$-axis are used, without twisted
periodic conditions. However, this inevitably increases the influence
of both the periodic boundary conditions and the finite number of
spheres on the search for the densest packings. We focus on slightly
elliptic cylinders with minor axis $D_{b}\in\left[1.6\textrm{,}2\right]$
and aspect ratios $D_{a}/D_{b}\in\left[1\textrm{,}1.10\right]$, and
employ the Monte Carlo (MC) method with simulated annealing for sphere
numbers $N\in\left[13\textrm{,}25\right]$. All cases with $N\leq12$
are well covered, since they are the factors of $N\in\left[13\textrm{,}25\right]$.

We employ the Monte Carlo method with simulated annealing, where particles
move according to the Metropolis algorithm. The overlap energies between
two spheres, $E_{ij}^{\textrm{S}}$, and between a sphere and the
wall, $E_{i}^{\textrm{B}}$, are defined as

\begin{equation}
E_{ij}^{\textrm{S}}=\begin{cases}
\frac{1}{2}(r_{ij}-d)^{2} & r_{ij}\leq d\\
0 & r_{ij}>d
\end{cases}\textrm{,}\label{eq:1}
\end{equation}

and

\begin{equation}
E_{i}^{\textrm{B}}=\begin{cases}
\frac{1}{2}\left(r_{i\textrm{B}}-\frac{d}{2}\right)^{2} & r_{i\textrm{B}}\leq\frac{d}{2}\\
0 & r_{i\textrm{B}}>\frac{d}{2}
\end{cases}\textrm{.}\label{eq:2}
\end{equation}

Here, $r_{ij}$ is the distance between the centers of two different
hard spheres, and $r_{i\textrm{B}}$ is the distance between a sphere's
center and the elliptic cylinder wall. It is worth noting that when
the total energy is zero, there is no overlap between the spheres
and between the spheres and the wall, and only then do the spheres
become truly hard spheres. By adjusting the cylinder height, we are
effectively searching for the case in which the spheres just separate
from each other and from the wall. At this point, the spheres become
hard spheres, and the resulting structure corresponds to a possible
densest packing structure \citep{Mughal2012}.

We consider a series of elliptic cylinders with various minor axes
$D_{b}$ and aspect ratios $D_{a}/D_{b}$ to obtain the densest packing
structure for each case. For a specific $D_{b}$ and $D_{a}/D_{b}$,
the simulation process is as follows:

1. For a given number of hard spheres $N$ and a cylinder length $L$,
simulated annealing is used to obtain the optimal solution (the minimum
total energy).

2. The length $L$ is then varied to find the minimum $L$ at which
the total energy is zero (spheres and wall are just separated), and
the packing fraction $\phi$ is computed.

3. The procedure is repeated for different $N$ to find the sphere
number corresponding to the maximum $\phi$.

4. Multiple runs (at least five per case) are performed, and the optimal
result is taken as the possible densest packing for the corresponding
$D_{b}$ and $D_{a}/D_{b}$.

In this work, the minor axis $D_{b}$ of the elliptic cylinder is
in the range $D_{b}\in\left[1.6\textrm{, }2\right]$. This range corresponds
to the zigzag, single-helix, and double-helix phases observed in the
circular cylinder (viewed as an elliptic cylinder with $D_{a}/D_{b}=1$,
where $D_{b}$ is the circular cylinder diameter $D$). An equal number
of experimental points are taken from each of the three phases to
avoid the issue of uniform sampling leading to too few points in certain
phases. For aspect ratios $D_{a}/D_{b}\in\left[1\textrm{, }1.10\right]$,
since the focus is on elliptic cylinders approaching the circular
cylindrical limit, more points are sampled when $D_{a}/D_{b}$ is
small and fewer when it is large; practically, uniformly selecting
points based on the eccentricity $e=\sqrt{1-\left(\frac{D_{b}}{D_{a}}\right)^{2}}$
fulfills this requirement. Further details are provided in Section
H of the Supplemental Material \citep{Supplemental}.

Based on the theory of packing in circular cylinders \citep{Chan2019},
the helix phases feature spheres with a coordination number of 4,
where every three spheres contact both the wall and one another, forming
triplets of spheres {[}Fig. \ref{fig:2}(a){]}. In the helical phases
of elliptic cylinders, triplets of spheres still exist. A polar coordinate
system in the $xy$-plane, with the ellipse’s center as the origin,
is used to specify sphere positions via the angular coordinate $\theta$
and the radial distance $r\left(\theta\right)$. Consider three spheres
positioned at $r_{i}\triangleq r\left(\theta_{i}\right)$, $r_{i+1}\triangleq r\left(\theta_{i+1}\right)$,
and $r_{i+2}\triangleq r\left(\theta_{i+2}\right)$, with angular
differences $\Delta\theta_{i\textrm{,}i+1}$, $\Delta\theta_{i+1\textrm{,}i+2}$,
and $\Delta\theta_{i\textrm{,}i+2}$. When these spheres form a triplet
of spheres with heights $z_{i}<z_{i+1}<z_{i+2}$ (from the top view,
a counterclockwise arrangement of spheres $i$, $i+1$, $i+2$ corresponds
to a right-handed triplet; otherwise, it is left-handed), the following
condition is satisfied (see derivation in Section B of the Supplemental
Material \citep{Supplemental}):

\begin{equation}
\begin{array}{c}
\sqrt{d^{2}-\left(r_{i}-r_{i+2}\right)^{2}-4r_{i}r_{i+2}\sin^{2}\frac{\Delta\theta_{i\textrm{,}i+2}}{2}}\\
=\sqrt{d^{2}-\left(r_{i}-r_{i+1}\right)^{2}-4r_{i}r_{i+1}\sin^{2}\frac{\Delta\theta_{i\textrm{,}i+1}}{2}}\\
+\sqrt{d^{2}-\left(r_{i+1}-r_{i+2}\right)^{2}-4r_{i+1}r_{i+2}\sin^{2}\frac{\Delta\theta_{i+1\textrm{,}i+2}}{2}}
\end{array}\textrm{.}\label{eq:1-1}
\end{equation}

Given $\theta_{i}$ and $\Delta\theta_{i\textrm{,}i+1}$, solving
the equation determines $\Delta\theta_{i+1\textrm{,}i+2}$, yielding
$\theta_{i+1}\triangleq\theta_{i}+\Delta\theta_{i\textrm{,}i+1}$
and $\theta_{i+2}\triangleq\theta_{i}+\Delta\theta_{i\textrm{,}i+1}+\Delta\theta_{i+1\textrm{,}i+2}$.
The vertical height differences $\Delta z_{i\textrm{,}i+1}$, $\Delta z_{i+1\textrm{,}i+2}$
and $\Delta z_{i\textrm{,}i+2}$ can then be calculated. By Eq. \ref{eq:1-1},
the positions of the $i$-th and $\left(i+1\right)$-th spheres entirely
determine the position of the $\left(i+2\right)$-th sphere, and subsequently
determine the packing of $N$ spheres. So we can vary the parameters
$\theta_{1}$ and $\Delta\theta_{1\textrm{,}2}$ to generate distinct
configurations and then seek out the one that yields the maximum packing
fraction. We call this approach the triplet-sphere packing block method.

\textit{Funding.}---
This work was financially supported by the National Natural Science
Foundation of China (12275137).

%


\end{document}